\documentclass[twocolumn,showpacs,preprintnumbers,amsmath,amssymb,pre,nofootinbib,superscriptaddress]{revtex4-1}
\usepackage[utf8]{inputenc}
\usepackage[T1]{fontenc}
\usepackage{multirow}
\usepackage{mathptmx}
\usepackage[export]{adjustbox}
\usepackage{mathtools}
\usepackage{float}
\usepackage{color}    
\usepackage{graphicx}
\usepackage{dcolumn}
\usepackage{bm}
\usepackage{subfigure}
\usepackage{amssymb}
\usepackage{amsmath}
\usepackage{float}
\graphicspath{{plots/}}

\usepackage{siunitx}
\DeclarePairedDelimiterXPP\BigOSI[2]%
  {\mathcal{O}}{(}{)}{}%
  {\SI{#1}{#2}}
  
\begin{document}

\title{Demarcating the classical and quantum approaches for the Coulomb logarithm in plasmas}

\author{S.K. Kodanova}
\affiliation{ Institute for Experimental and Theoretical Physics, Al-Farabi Kazakh National University, 71 Al-Farabi ave., 050040 Almaty, Kazakhstan}
\affiliation{Institute of Applied Sciences and IT, 40-48 Shashkin str., 050038 Almaty, Kazakhstan}

\author{T.S. Ramazanov}
\affiliation{ Institute for Experimental and Theoretical Physics, Al-Farabi Kazakh National University, 71 Al-Farabi ave., 050040 Almaty, Kazakhstan}
\affiliation{Institute of Applied Sciences and IT, 40-48 Shashkin str., 050038 Almaty, Kazakhstan}

\author{M.K. Issanova}
\email{issanova@physics.kz}
\affiliation{ Institute for Experimental and Theoretical Physics, Al-Farabi Kazakh National University, 71 Al-Farabi ave., 050040 Almaty, Kazakhstan}
\affiliation{Institute of Applied Sciences and IT, 40-48 Shashkin str., 050038 Almaty, Kazakhstan}

%\date{\today}

\begin{abstract}
The Coulomb logarithm often enters various plasma models and simulation methods for computing the transport and relaxation properties of plasmas. Traditionally, a classical pair collision picture was used to calculate the Coulomb logarithm for different plasma parameters. 
With the recent emergence of the high interest in dense plasmas with partially degenerate electrons, new approaches have been developed to treat electron-ion collisions in a quantum-mechanical way.
In this context, gaining a deeper physical understanding of the criteria for the applicability of classical plasma models is crucial. 
We have analyzed the Coulomb logarithm describing the electron-ion collisions in hydrogen plasmas in a wide range of temperatures and densities relevant to inertial confinement fusion experiments. The electron-ion collision cross-sections were computed using both quantum and classical scattering theories. We found that the classical description of the electron-ion scattering in plasmas and related plasma models is applicable if the de Broglie wavelength of electrons is smaller than the ion charge screening length in plasmas. Additionally, we show that the quantum first-order Born approximation for describing the electron-ion collision is accurate only in the regime where classical scattering theory is accurate. 

\end{abstract}

\maketitle

\section{\label{Sec. I} Introduction}

Interest in high energy-density plasmas is fueled by the ongoing work on harnessing inertial confinement fusion (ICF). In the ICF experiments, the plasma state passes the regime of partial degeneracy with a high density of particles. This plasma state is often called warm dense matter (WDM). 
Accordingly, understanding and computing plasma transport properties across temperature and density regimes is one of the main topics in the field of high-energy-density plasmas.  In this regard, 
the kinetic theory-based plasma models that utilize scattering cross-sections and related Coulomb logarithms are often used to compute plasma transport properties. The Coulomb logarithm emerges naturally in the collision integral occurring in the kinetic equations.
Therefore, the Coulomb logarithm is one of the key parameters in various plasma models and, as a result, plays a decisive role in studying the transport and dynamic properties of plasma state\cite{Filippov}.
For example, adequate computing of the Coulomb logarithm is important for the modeling of plasmas in stellar envelopes \cite{Daligault}, dusty plasma \cite{Khrapak, Khrapak2, Kodanova_pop_2015}, ultracold plasma \cite{Castro}, inertial confinement fusion \cite{Sprenkle, Vorberger1, Dharma-wardana, Vorberger2}, and to model laser ablation \cite{Skupsky}.

In classical plasma models, the divergence in the collision integral at small angles is eliminated by effective screening of pair interaction potential of plasma particles \cite{Spitzer, Moldabekov_pop_2015}.
The occurrence of such divergence due to scattering at small angles $\theta_{min}\ll1$ can be seen by considering the classical  momentum transfer cross section of two point-like charges $q_1$ and $q_2$ colliding with the initial relative velocity $\upsilon$:
\begin{equation} \label{Qum} 
Q_{m}= 4\pi \left[\frac{q_{1}q_{2}}{m_{12}\upsilon^{2}}\right]^{2} \ln \left[b_{max}\frac{m_{12}\upsilon^{2}}{|q_{1}q_{2}|}\right],  
\end{equation} 
where $m_{12}^{-1}=m_{1}^{-1}+m_{2}^{-1}$ defines the reduced mass of colliding particles and $b_{max}$ is the largest value of the impact parameter in classical scattering of two particles. The  $b_{max}$ value corresponds to the minimum scattering angle $\theta_{min}$. The logarithmic factor in Eq.(\ref{Qum}) is called Coulomb logarithm \cite{Spitzer}:
\begin{equation} \label{Coulomblog} 
{\Xi} =\ln \frac{b_{max}}{b_{min}} .
\end{equation} 

In fact, most of the used plasma models describing energy transfer contain a Coulomb logarithm similar to Eq. (\ref{Coulomblog}) with certain modifications depending on the type of colliding particles and used pair-interaction potential \cite{Baalrud, MRE18, Moldabekov_cpp_2022}. The cutoff parameter $b_{\rm max}$ is usually set to screening length (Debye length), and the particles are considered not to interact at distances larger than screening length. In a dense medium, pair-collision approximation becomes inaccurate, and the effect of the presence of other particles on scattering has to be taken into account \cite{Starrett_pre_2012, Daligault_prl_2016}, going beyond the mean-field screening level.  

Elimination of the divergence at the lower limit of a Coulomb logarithm (large scattering angles) is also carried out by choosing a finite value of $b_{min}$ (cf. Eq.(\ref{Coulomblog})) that corresponds to a physically reasonable mechanism at close pair collisions. In a classical picture, $b_{min}$ is chosen to be the classical distance of the closest approach $r_0$. For partially degenerate electrons, $b_{min}$ is often set to the electron thermal de Broglie wavelength.
% $\lambdabar_{\alpha \beta}$ (with $\alpha$ and $\beta$ denoting the types of colliding particles).

\begin{figure}[t]
\includegraphics[width=\linewidth]{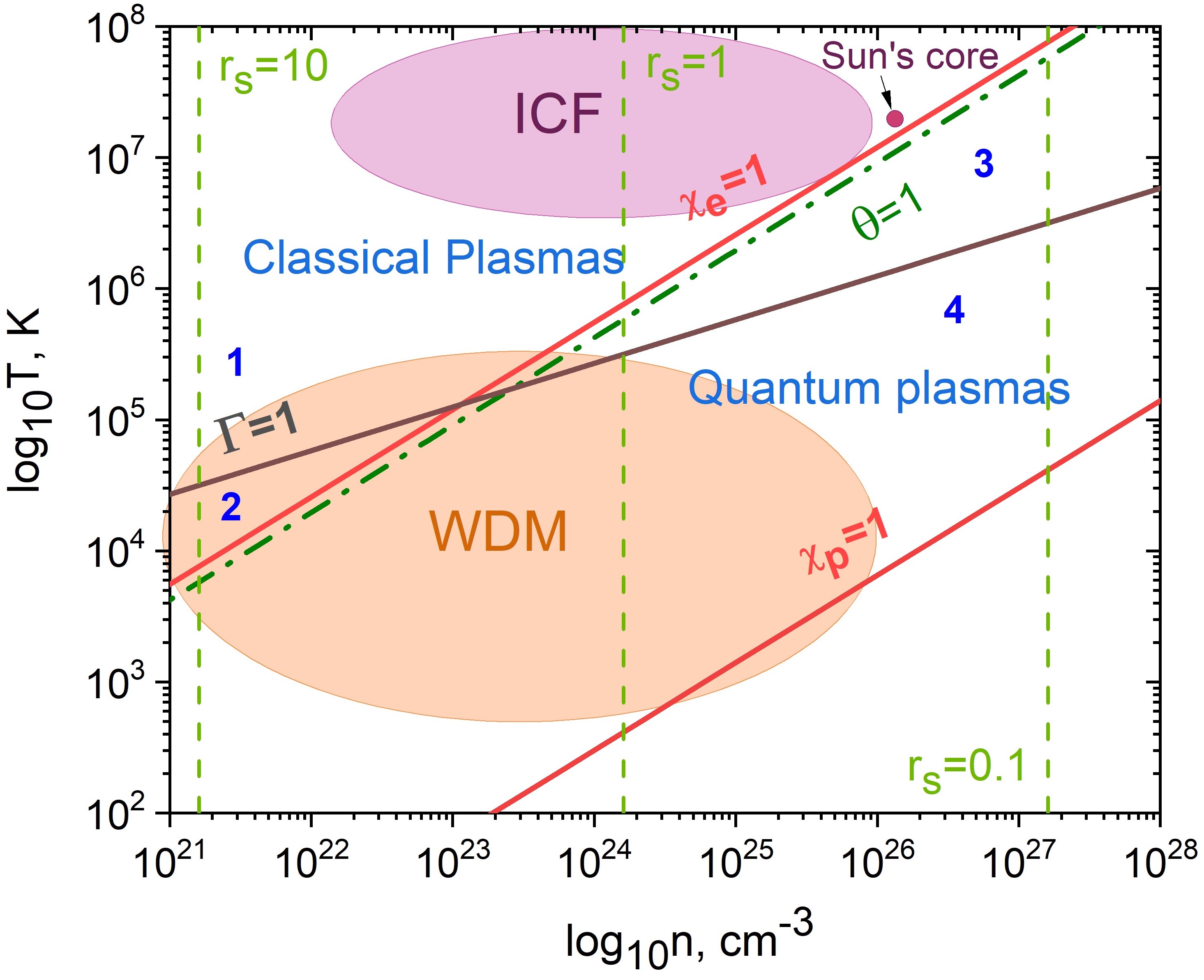}
\caption{Parameter regimes of dense plasmas. The
solid black line is the boundary between weak and strong electron
coupling $\Gamma=1$; the dash-doted line approximately separates plasma parameters corresponding to classical and degenerate electrons $\theta=1$; the orange oval denotes the parameters corresponding to warm dense matter regime; the purple oval denotes the parameters relevant for ICF plasmas. The dashed light green lines indicate different values of $r_{s}$.}
\label{fig01}
\end{figure}

Depending on plasma parameters, methods with different levels of complexity and with different approximations are used to compute plasma transport coefficients. 
In Fig. \ref{fig01}, we depict the temperature and density parameters related to  WDM and ICF plasma. In Fig.\ref{fig01}, we also show the temperatures and density at which characteristic thermal energy $k_BT$ is comparable to the quantum Fermi energy of electrons. This partial degeneracy is depicted by the $\theta=k_BT/E_F=1$ line in Fig.\ref{fig01}. In addition, in Fig.\ref{fig01}, we show the temperatures and densities at which the characteristic  Coulomb interaction energy between protons $U_C=e^2/a$ (with $a$ being the mean interparticle distance) is comparable with the thermal energy of protons $k_BT$. This regime is characterized by the coupling parameter $\Gamma=U_C/k_BT \gtrsim 1$.
% The parameter range defined by $\Gamma=1$ and $\theta=1$ delineates distinct regions in density-temperature characteristics. 
In Fig.\ref{fig01}, we distinguish (1) semi-classical weakly coupled regime, (2) semi-classical strongly coupled regime, (3) quantum weakly coupled state, and (4) plasmas with strongly coupled ions and with degenerate weak or strongly coupled electrons (the domain below $\Gamma=1$ and $\theta=1$ lines). Typical parameters of WDM  depicted by the orange oval and the purple oval correspond to the domain of ICF plasma.
Transport phenomena in the left corner of region (1) are well described by Landau-Spitzer theory \cite{Spitzer}, whereas quantum weak-coupling theories, such as the quantum Landau-Fokker-Planck equation \cite{Daligault16}, have been effective in modeling region (3). Recently,  classical plasma transport theory was extended to domain (2) using so-called mean force kinetic theory (MFKT) \cite{Daligault13}, which has also been successful in domain (4) for modeling ion transport in WDM regime \cite{Starrett16}. Furthermore, various kinetic approaches, including binary collision theories \cite{Gericke}, linear response theories \cite{Scullard, Daligault09}, and nonequilibrium Green’s functions along with field-theoretic methods \cite{Balzer, Bonitz}, are employed to compute transport coefficients in WDM.

Additionally, for plasmas with non-ideal ions $\Gamma>1$, in Ref. \cite{Baalrud}, it was shown that one can calculate the transport properties of ions using the Chapman-Enskog method for the Coulomb logarithm, where the classical scattering in the field of an effective pair interaction potentials defined in terms of the logarithm of the radial pair correlation function is considered. 
For two-component weakly degenerate plasmas, among existing models for the Coulomb logarithm, we mention often-used approximations by Gericke, Murillo, and Schlanges (GMS) \cite{Gericke}, by Brown, Preston, and Singleton (BPS) \cite{Brown}, and by Brown and Singleton \cite{Brown2007}.
Furthermore, Ramazanov and coworkers have investigated transport properties of two-component plasmas using the method of effective potentials with screening naturally included using linear response theory \cite{Kodanova1, Kodanova2, Ramazanov_pre_2016, Moldabekov1, Moldabekov19, Kodanova3, Moldabekov_cpp_2017, Issanova_2016}.

Depending on plasma parameters, classical or quantum scattering cross-sections are used to compute the transport properties of plasmas.
In practice, it is helpful to know when classical plasma models can be used and quantum effects are negligible. The parameters at which quantum effects are relevant depend on the type of colliding particles involved in characterizing a given transport coefficient. In this work we focus on the ion-electron collision. We compute transport cross-sections and the Coulomb logarithm using both quantum and classical scattering theories. 
Considering different temperatures and densities, we determine plasma parameters at which a classical description is highly accurate. 
The presented analysis provides insight into when a standard classical plasma picture is applicable and when quantum theory must be used. This is important, e.g., for interpreting plasma experiments. In addition, this study deepens our understanding of the dense plasma state.  

In Sec. \ref{sec: I}, we introduce the generalized Coulom logarithm in terms of the electron-ion transport cross-section and show the used quantum and classical calculation methods. The result of the calculations and corresponding discussions are provided in Sec. \ref{Sec. II}.
We conclude the paper by summarizing our main findings.

\section{\label{sec: I} Theoretical Model}

\subsection{Generalized Coulomb logarithm}

% In many works, the transport cross section of particle scattering in the quantum-mechanical approximation based on the calculation of scattering phases with different interaction potentials \cite{POP2021, CPP2024}. Another effect associated with the replacement of Maxwell-Boltzmann statistics by Fermi-Dirac statistics in the quantum approximation, i.e. at high densities or at low temperatures, which can be found in the work \cite{Rightley}.

% In this regard, it is important to show the criterion of applicability of various approximations used in calculating scattering cross sections. Here we limited ourselves to considering electron-ion contributions to transfer processes, and taking into account the contribution of electron-electron interaction will be the next step both in theory and in the development of modeling. We will consider when it is necessary to take into account quantum-mechanical effects in scattering and their influence on collision integrals and, accordingly, on the Coulomb logarithm calculated on the basis of these $\Omega$-integrals. Comparisons of these results are carried out and the degeneracy parameters and density parameters are shown at which the value of the Coulomb logarithm exactly coincides with each other within $1\%$ errors. Calculations are carried out using classical theory and quantum-mechanical theory.

We consider the generalized Coulomb logarithm
determining such important transport characteristics as plasma conductivity and the electron-ion relaxation rate \cite{Rightley, Daligault_pre_2019}.  Accordingly, the generalized Coulomb logarithm is computed using the transport cross-section $Q^{T}(g)$ defined by electron-ion collisions \cite{Rightley}:
\begin{equation} \label{Xi}
\Xi= {\frac{1}{2} \int_{0}^{\infty} dg G(g) \frac{Q^{T}(g)}{\sigma_{0}}},
\end{equation}
where $\sigma_{0} =\hbar ^{2} /(m_{e} v_{\rm th})^{2}$, $g=u/v_{\rm th}$, $u$ is the relative velocity of the scattering particles, and $v_{\rm th}=\sqrt{2k_BT/m_{ei}}$. 

In Eq. (\ref{Xi}), the function $G(g)$ determines the relative availability of states that contribute to the scattering \cite{Rightley}. The factor $G(g)$ depends on the statistics used for ions and electrons.
The transport cross-section $Q^{T}(g)$ and $G(g)$ can be computed using a quantum or classical approach. We use both methods and compare the results to evaluate the impact of quantum effects.

\subsection{Quantum description of the Coulomb logarithm}

Typically, plasma physics primarily considers high-temperature, low-density regimes for which quantum effects are negligible. 
However, with increasing density, the average distance between particles decreases, and their quantum nature can no longer be ignored. Consequently, it is often assumed that the quantum effects are relevant for plasma when the thermal wavelength of electrons ${\lambdabar}_{\alpha \beta } $ becomes comparable to the average interparticle distance (see the $\chi_e=1$ line in Fig.\ref{fig01}), where
the thermal de Broglie wavelength for a pair of particles $\alpha $ and $\beta $ is determined by the characteristic velocity of a thermal motion of particles $v_{\rm th}$ and reduced mass of colliding particles $m_{\alpha \beta } ={m_{\alpha } m_{\beta }/ (m_{\alpha } +m_{\beta }) }$ as $\lambdabar_{\alpha \beta} =\hbar/(m_{\alpha \beta } v_{\rm th})$. 

Statistical weighting function $G(g)$ derived using Fermi-Dirac distribution for electrons and Maxwell distribution for ions reads \cite{Rightley}
\begin{equation} \label {functionG}
G(g)= {\frac{\eta \exp^{-g^{2}}g^{5}} {[-Li_{3/2}(-\eta)](\eta e^{-g^{2}}+1)^{2}}},
\end{equation}
where $-{\rm Li}_{3/2}=[-\eta]=\frac{4}{3\sqrt{\pi}} \theta^{-3/2}$, $\eta=\exp(\mu/k_{B}T_e)$, and $\mu$ is the electron chemical potential \cite{Melrose}.

Within quantum scattering theory, the  transport cross-section $Q^{T}$ is computed as ~\cite{Landau}:

\begin{align}
\label{Transport}
Q^{T} (k)&=\frac{4\pi }{k^{2} } \sum _{l}(l+1)\sin ^{2}  (\delta _{l} (k)-\delta _{l+1} (k)),
\end{align}
where $k=u\,m_{ei} /\hbar = g v_{\rm th} m_{ei}/\hbar$ is the scattering wave number.

The phase shifts $\delta _{l} (k)\equiv \delta _{l} (k,r\to \infty )$ in Eq. (\ref{Transport}) are computed by solving the Calogero equation \cite{Calogero, Babikov, Babikov0}: 
\begin{equation} \label{Phase} 
\frac{d}{dr} \delta \left(r\right)=-\frac{\Phi (r)}{k} \left[\cos \delta _{l} (r)j_{l} (kr)-\sin \delta _{l} (r)n_{l} (kr)\right]^{2} ,
\end{equation}
with the condition $\delta _{l} \left(0\right)=0.$ In Eq.~(\ref{Phase}), $k$ is the scattering wave number, $l$ indicates the orbital quantum number, $j_{l}$ and $n_{l} $ are the Rikkati-Bessel functions, and $\Phi(r)$ is the pair interaction potential.

% As is known, in a quantum mechanical calculation for each $g$ it is necessary to calculate the phase shift $\delta _{l} (r)$ for different values of the angular momentum $l$ based on the Calogero equation (\ref{Phase}) instead of finding the scattering angle for each value impact parameter $b$.

To analyze and cross-check the results, it is useful to compare the data obtained from the exact solution of the  Calogero equation with the results computed using the first-order Born approximation.
The first-order Born approximation is exact in the limit of large wavenumbers (energies), where the field of scattering potential is weak and $\delta _{l} (k)\ll \pi/2$. Thus, this can be considered a valuable test of the numerical solution of Eq. (\ref{Phase}) at high energies.   The phase shifts in the first-order Born approximation are computed using the following well-known result~\cite{Babikov, Babikov0}:

\begin{equation} \label{Born} 
\tan \left[\delta_{l}(r,k)\right]=-\frac{1}{k} \int _{0}^{r}\Phi (r')j_{l}^{2}  (kr')dr'.
\end{equation}

The comparison of the Coulomb logarithm computed using the exact solution for pair quantum scattering with the results from the first-order Born approximation allows us to identify the plasma parameters where electron-ion coupling can be considered weak on average.

\subsection{Classical description of the Coulomb logarithm}

In the classical Chapman-Enskog theory, 
statistical weighting function $G(g)$ follows from using the Maxwell distribution for both electrons and ions \cite{Chapman, Stanton, Daligault13, Baalrud_Anizotropy}:
\begin{equation} \label {functionG}
G(g)= e^{-g^2}g^{5}.
\end{equation}

% generalized  Coulomb logarithm associated with the  collision integrals :

%\begin{equation} \label{Eq.4} 
%\Omega ^{(l,r)} =\sqrt{\frac{k_{B} T}{2\pi m_{ij} } } \int _{0}^{\infty }\exp (-g^{2} ) \, g^{2r+3} Q_{ij}^{(l)}(g)dg,        
%\end{equation}

% \begin{equation} \label{Kul_cl} 
%  \Xi_{cl}=\frac{1}{2} \int _{0}^{\infty } dg e^{-g^2} \, g^{5}  \frac{Q_{ij}^{(l)}(g)}{\sigma_{0}},
% \end{equation}
% where $m_{ij}=m_im_j/(m_i+m)j) $ is the reduced mass of colliding particles of types $i$ and $j$, and $Q_{ij}^{(l)}(g) $ is the momentum-transfer cross-section particles colliding with a relative reduced velocity $g=v/v_{\rm th}$ with $v_{\rm th}=\sqrt{2k_BT/m_{ij}}$.

The classical momentum-transfer cross-section for the considered case is computed using the scattering angle of the binary collision as: 
\begin{equation} \label{Eq.5} 
Q^{T}(g) =2\pi \int _{}^{}\left(1-\cos \chi \left(b,g\right)\right) bdb.
\end{equation}
where $b$ is the impact parameter of the pair collision and the scattering angle $\chi(b,g)$ is determined by the relation
\begin{equation} \label{Eq.7} 
\chi(b,g) =\pi -2b\int _{r_{0} }^{\infty }\frac{\mathrm{d}r\,r^{-2}}{\sqrt{1-2\Phi(r)/(m_{ei}v^2_{\rm th} g^{2}) -b^{2}/r^{2}} } .         
\end{equation}

Eq. (11) is a standard approach for the calculation of the various transport characteristics in classical plasmas \cite{Khrapak, Stanton}

\subsection{Pair interaction potential}
We consider hydrogen plasma and for electron-ion pair interaction potential we use a Debye-type form of the screened  potential:
\begin{equation} \label{Eq.2} 
\Phi (r)=\frac{e^{2} }{r} \exp({-r/\lambda_s}),        
\end{equation}
where $\lambda_s$  is the screening length.

The screening length in electron-proton plasmas is computed as
\begin{equation}\label{eq:lambda_s}
    \lambda_s=\left(\frac{1}{\lambda_e^2}+\frac{1}{\lambda_i^2+a_i^2}\right)^{-1/2},
\end{equation}
where $\lambda_e$ is the electron screening length,  $\lambda_i=\left(k_BT/(4\pi n_ie^2)\right)^{1/2}$ is the Debye screening length of ions, and $a_i$ is the mean distance between ions. 
Stanton and  Murillo \cite{Stanton} have demonstrated that using  Eq. (\ref{eq:lambda_s}) provides an accurate description of the transport properties of ions across coupling regimes. 

\begin{figure}[t]
\includegraphics[width=0.905\linewidth]{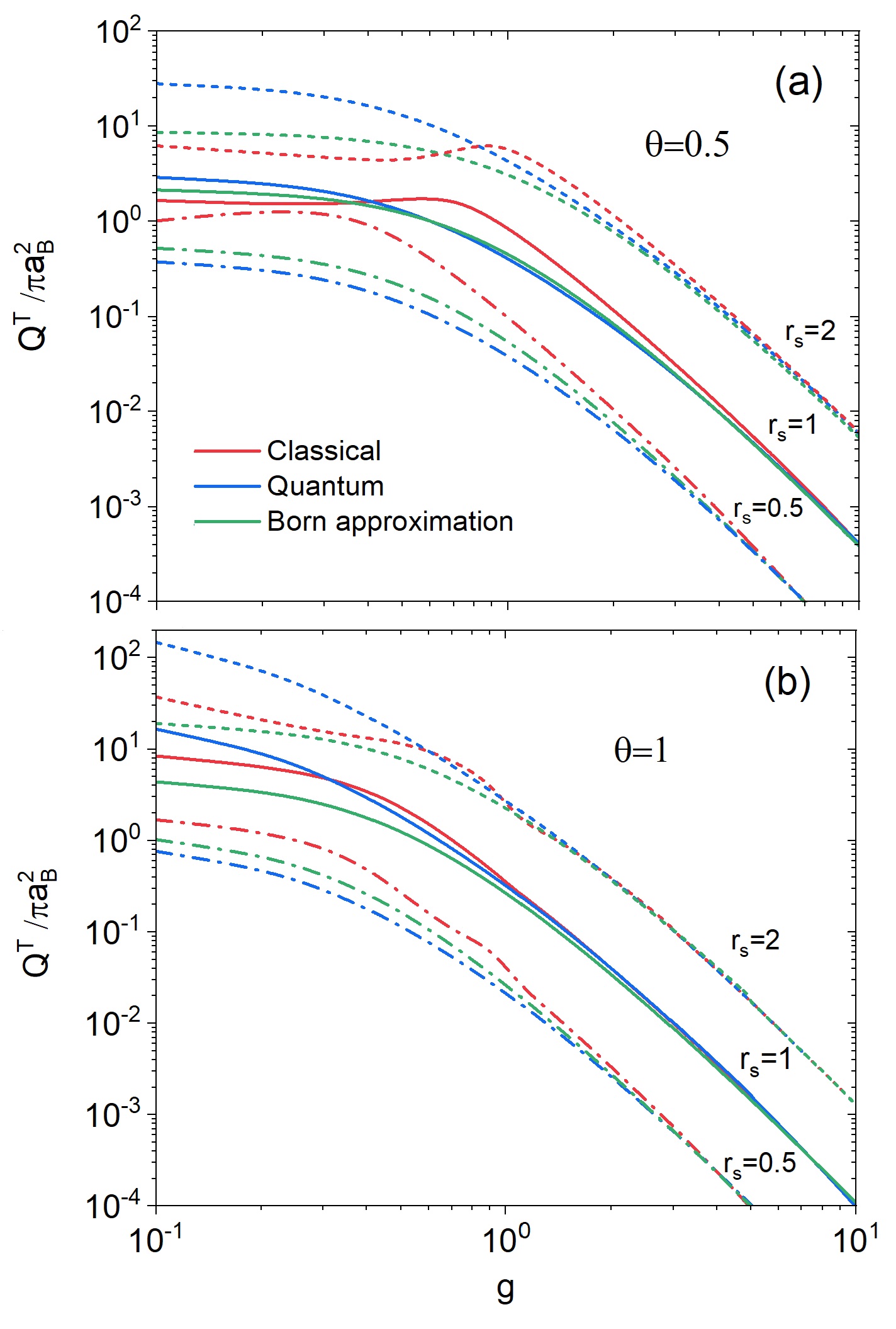}
\caption{The transport cross-sections at different values of $r_{s}$. Panel (a) shows the results for the degeneracy parameter $\theta=0.5$, and panel (b) shows the data at the degeneracy parameter $\theta=1$.}
\label{fig02}
\end{figure}

For the electron screening length, we use the general result from the long wavelength limit of the density response function in random phase approximation \cite{PhysRevA.29.1471, Moldabekov_pop_2015, Moldabekov_cpp_2017}:
\begin{equation}
    \lambda_e^{-2}=\lambda_{TF}^{-2} \cdot \theta ^{1/2} \cdot I_{-1/2} \left(\eta \right)/2,
\end{equation}
where $\lambda_{TF}=v_F/(\sqrt{3}\omega_{p})$ is the Thomas-Fermi screening lenght at $T\to0$, $v_F$ is the Fermi velocity, $\omega_{p}$ is the electron plasma frequency,  and 
$I_{-1/2} \left(\eta \right)$ is the Fermi integral of order $-1/2$ at the reduced chemical potential of electrons $\eta=\mu/(k_BT)$.

To describe the plasma state, we use the dimensionless density parameter $r_{s} =a/a_B$ (defined as the ratio of the mean interelectronic distance to the first Bohr radius) and electron degeneracy parameter $\theta =\frac{k_{B} T}{E_{F} }$. Different $r_s$ values are also depicted in Fig. \ref{fig01}. We consider dense hydrogen plasma with $r_{s} \leq 4$ and $\theta\leq 5$.

\begin{figure}[t]
\includegraphics[width=0.9\linewidth]{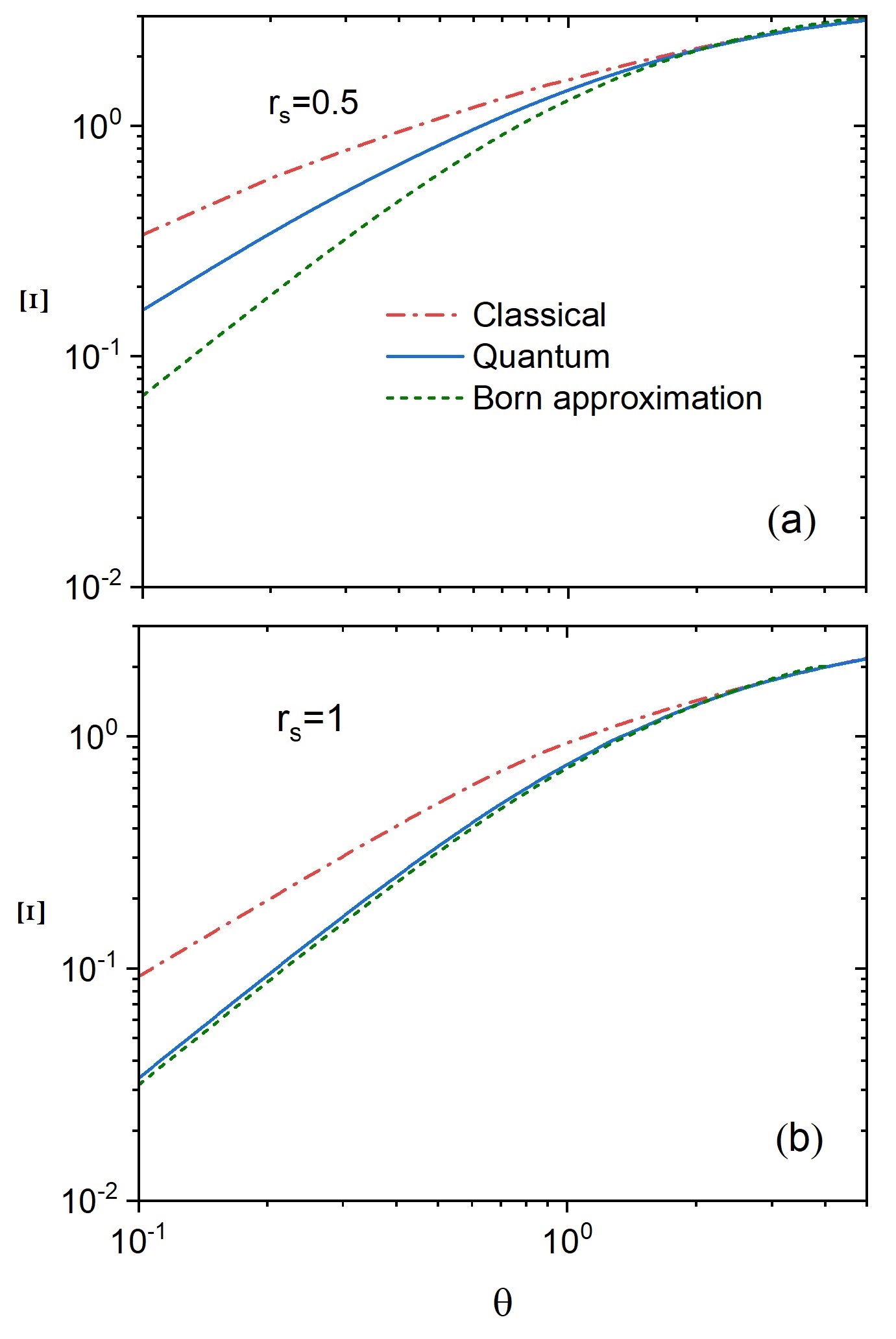}
\caption{The Coulomb logarithm (\ref{Xi}) computed using quantum and classical methods for different degeneracy parameters at (a) $r_{s}=0.5$ and at (b) $r_{s}=1$.}
\label{fig03}
\end{figure}

\begin{figure}[t]
\includegraphics[width=0.9\linewidth]{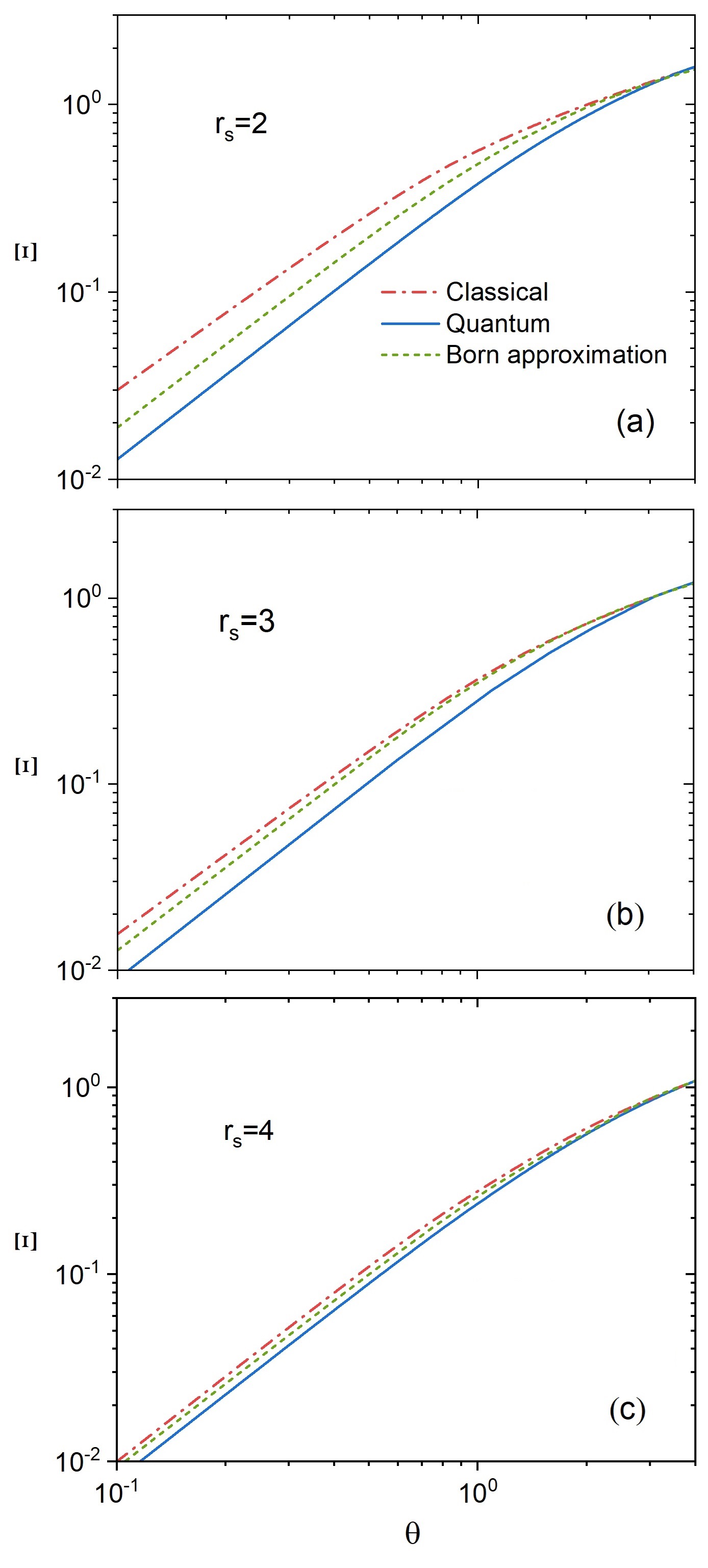}
\caption{The Coulomb logarithm (\ref{Xi}) computed using quantum and classical methods for different degeneracy parameters at (a) $r_{s}=2$, at (b) $r_{s}=3$, and at (c) $r_{s}=4$.}
\label{fig04}
\end{figure}

\begin{figure}[t]
\includegraphics[width=\linewidth]{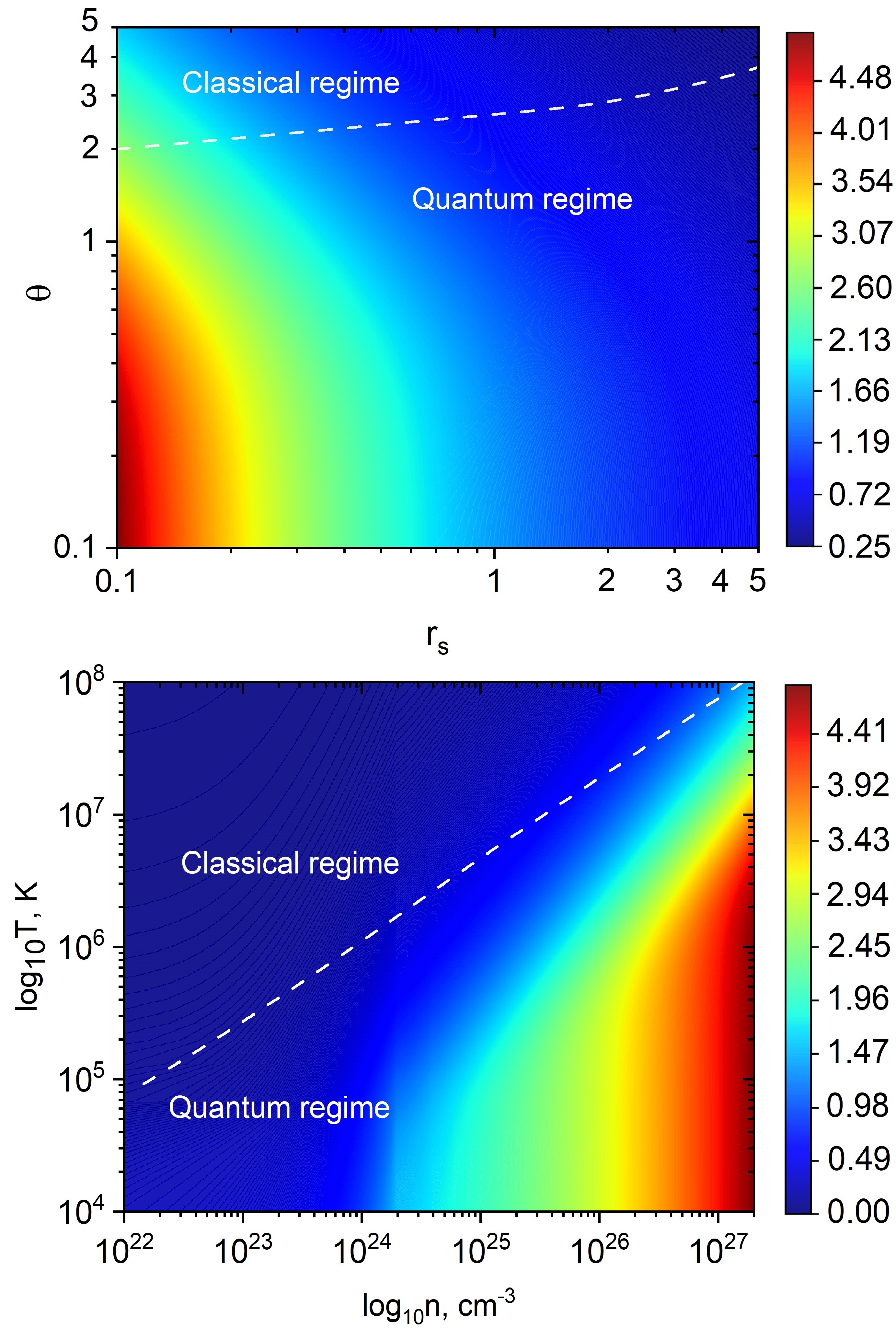}
\caption{The plasma parameters at which classical picture for the electron-ion collision is accurate are in the domain above dashed line. The color map shows the ratio of the electron de Broglie wavelength to the plasma screening length. The upper panel presents results on the $\theta$-$r_s$ plane, and the lower panel corresponds to the $T$-$n$ plane.}
\label{fig05}
\end{figure}

\section{\label{Sec. II} Results and discussions}

In Fig. \ref{fig02}, we show the results for the transport cross-section computed for $\theta=0.5$ (panel (a)) and $\theta=1.0$ (panel (b)) with the density parameter values $r_s=0.5$, $r_s=1$, and $r_s=2$.
From Fig. \ref{fig02}, we see that the scattering cross-section increases with the decrease in the relative velocity of colliding particles. This is due to stronger coupling between colliding particles at lower collision velocities.  The classical cross-section shows a non-monotonic behavior
at $u<v_{\rm th}$ (e.g., it is clearly visible at $\theta=0.5$ and $r_s=2$); this is due to the emergence of quasi-orbital motion of particles during the collision process \cite{Stanton, Khrapak, Kodanova_pop_2015}. In contrast, the quantum scattering cross-section does not have such non-monotonic behavior at small collision velocities. Here, we recall that at smaller momenta, the electron is stronger spread out in space (i.e., has a larger de Broglie wavelength), and the classical picture becomes unsuitable. Indeed, we see that the results for the classical cross-section is in good agreement with the quantum cross-section at large collision velocities with $u\gg v_{\rm th}$. The quantum and classical results start to substantially deviate from each other at $u\lesssim v_{\rm th}$. Depending on $\theta$ and $r_s$ values, the classical approach can overestimate or underestimate the scattering cross-section compared to the quantum collision calculations. As expected, the first-order Born approximation agrees with the exact solution of the quantum scattering problem at $u\gg v_{\rm th}$ and becomes invalid at low collision momenta.  
An interesting observation is that the deviation of the Born approximation from the exact quantum solution occurs at collision velocities where the classical solution also deviates from the quantum scattering cross-section. 
Furthermore, at high velocities, where the Born approximation is valid,  the classical cross-section agrees well with the first-order Born approximation results. 

Next, in Fig. \ref{fig03}, we show the result for the generalized Coulomb logarithm (\ref{Xi}) computed for $r_s=0.5$ and $r_s=1.0$ with the degeneracy parameter in the range $0.1\leq \theta \leq 5$.
From Fig. \ref{fig03}, we see that the classical Coulomb logarithm is in agreement with the quantum Coulomb logarithm at $\theta\gtrsim 2$. This is also the case at $r_s=2$, $r_s=3$, and $r_s=4$ as one can see from Fig. \ref{fig04}. The quantum effects in electron-proton scattering can not be neglected at lower degeneracy parameter values $\theta<2$. At $\theta\gtrsim 2$ and all considered $r_s$ values in Fig. \ref{fig03} and Fig. \ref{fig04}, the first-order Born approximation approximation is in agreement with the quantum and classical Coulomb logarithms that are computed solving scattering equations exactly. The first-order Born approximation provides the Coulomb logarithm values close to the exact quantum solution at $r_s=1$. However, this seems to be a coincidence.
Indeed, from Fig. \ref{fig03} and Fig. \ref{fig04}, we see that at $4\leq r_s \leq 0.5$, the values of the Coulomb logarithm computed using the first-order Born approximation relative to the exact quantum solution decreases with the decrease in $r_s$ and the agreement between two data sets appear only at $r_s=1$, but not at $r_s=0.5$. 

Finally, to provide a more accurate picture of where the classical scattering approach is valid, we show in Fig. \ref{fig05}(a) the domain of $\theta$ and $r_s$ values and in Fig. \ref{fig05}(b) the domain of temperatures $T$ and free electron densities $n$ where the generalized Coulomb logarithm computed using the classical scattering cross-section exactly coincides with the exact quantum solution within an error of $1\%$. 
According to this criteria, a dashed line in Fig. \ref{fig05} separates the classical and quantum scattering domains.
In addition, the color map in Fig. \ref{fig05} indicates values of the ratio of the de Broglie wavelength to the screening length $\lambdabar_{ei}(r_{s},\theta )/\lambdabar_{s}(r_{s},\theta)$. The main conclusion we draw from Fig. \ref{fig05} is that the classical picture for the election-ion collision can be used only when the characteristic de Broglie wavelength of an electron in plasma is smaller than the plasma screening length. 

\section{Conclusion}

We have analyzed the Coulom logarithm in dense hydrogen plasmas across temperature and density regimes. The calculations were performed using the exact solution of the scattering problem for electron-ion collision in quantum and classical pictures. In addition, we have analyzed the applicability of the first-order Born approximation. As the main result, we show the temperature and density values at which the classical models for the Coulomb logarithm are applicable. We show that the classical electron-ion scattering picture can be accurate only if the de Broglie wavelength of electrons is smaller than the charge screening length in plasmas.
This conclusion is also expected to remain valid for plasmas with different compositions. Another interesting result is that the quantum first-order Born approximation is accurate only at parameters where quantum effects in electron-ion scattering are negligible. 

\section*{Acknowledgments}
This research is funded by the Science Committee of the Ministry of Education and Science of the Republic of Kazakhstan Grant AP19677200 "Investigation of structural and dynamic properties of non-ideal plasma".

\bibliography{Main}
 
\end{document}